# Thermal soaring flight of birds and UAVs

Zsuzsa Ákos[1], Máté Nagy[1], Severin Leven[2] and Tamás Vicsek[1,3]

[1]Department of Biological Physics, Eötvös University, Pázmány Péter sétány 1A, H-1117, Budapest, Hungary,
[2]Laboratory of Intelligent Systems, École Polytechnique Fédérale de Lausanne, Switzerland,
[3]Statistical and Biological Physics Research Group of HAS, Pázmány Péter sétány 1A, H-1117, Budapest, Hungary.

Email: vicsek@hal.elte.hu

**Abstract.** Thermal soaring saves much energy, but flying large distances in this form represents a great challenge for birds, people and Unmanned Aerial Vehicles (UAVs). The solution is to make use of so-called thermals, which are localized, warmer regions in the atmosphere moving upwards with a speed exceeding the descent rate of birds and planes. Saving energy by exploiting the environment more efficiently is an important possibility for autonomous UAVs as well. Successful control strategies have been developed recently for UAVs in simulations and in real applications. This paper first presents an overview of our knowledge of the soaring flight and strategy of birds, followed by a discussion of control strategies that have been developed for soaring UAVs both in simulations and applications on real platforms. To improve the accuracy of simulation of thermal exploitation strategies we propose a method to take into account the effect of turbulence. Finally we propose a new GPS independent control strategy for exploiting thermal updraft.

**1. Introduction, history**

Thermal soaring is a form of flight where the flying objects use only convection currents, called thermals, to stay in the air without any additional power source (motor power in the case of airplanes or flapping of wings in the case of birds). Thermals are spatially and temporally localized parts of the atmosphere created by solar radiation heating the ground, typically moving upwards with a speed in the range of 1–10m/s. The ground heats up the air nearby which rises in columns. Man made soaring objects, like gliders, hang gliders and paragliders are able to fly great distances by using only the natural energy of thermals but larger species of birds have also specialized during evolution for this form of flight. Similarly to gliders these birds gain height by circling in thermals with wings spread until the desired height is reached. Then a more or less straight advancing but sinking phase follows until the next thermal is reached (Figure 1). This paper does not discuss other types of soaring, such as dynamic soaring, in which the energy for flying great distances is gained from horizontal wind gradients. Therefore we will refer to thermal soaring as soaring henceforth.

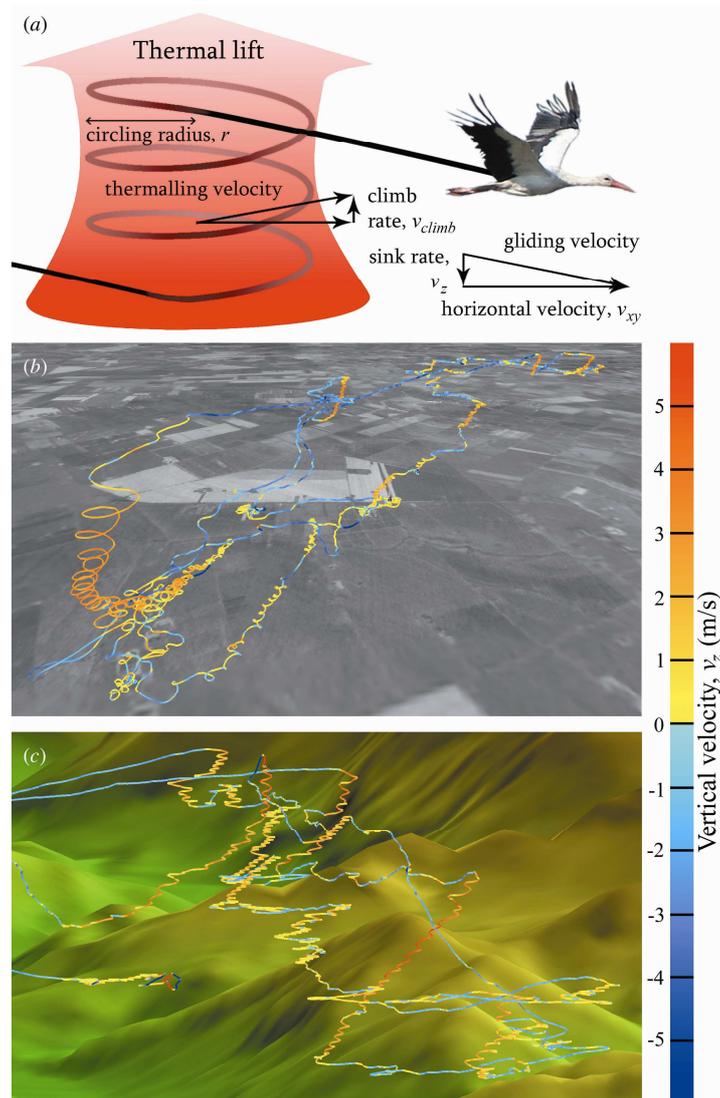

**Figure 1.** (*a*) Illustration of soaring flight with the notations indicated. (*b*) GPS logged flight path of a peregrine falcon (*Falco peregrinus*) on a black and white satellite map of the region. Colour scale indicates the vertical velocity, red corresponds to climbing (usually within thermals), blue to sinking (gliding). (*c*) GPS logged flight path of a paraglider with the local ground relief. Courtesy of Akos et al. [9]. Reproduced with permission, © 2008 PNAS.

Until about 1870 the soaring of birds was a mystery. It wasn't understood how birds are able to fly without flapping their wings. An explanation based on common sense suggested that they flew with slow flapping that could not be seen. Later, a scientist, I. Lanchaster, who spent 5 years investigating the soaring flight of birds, published some of his observations and deductions [1]. Consequently, soaring flight became an object of great interest in the scientific community.

Otto Lilienthal was the first who made systematic analysis of the problem. He observed sea birds (following vessels at sea), and storks, whose outstanding soaring ability is well known. He was the first who proposed that artificial wings should be designed with sustaining surfaces of concavo-convex shape. He and his brother, Gustav Lilienthal, carried out flying experiments for nearly 25 years, verifying their theory. He published a book titled "Bird Flight as the Basis of Aviation" in 1889 [2].

Soaring flight went through great development after World War I in Germany when the manufacture of powered airplanes was restricted by the Allies. Gliding soon became a highly favoured sporting activity among people who dreamed about using only nature's power like birds. Later, simpler and more popular directions of soaring flights developed, like hang gliding and recently paragliding, attracting more and more people.

**2. Research on soaring flight of birds**

After the theory of gliding flight was described research on soaring birds focused on the performance of birds' wings and the effect of the morphing wing on performance. The dependence of circling radius in thermals and boundary layer usage on wing parameters was also a subject of investigations.

*2.1 The glide polar (polar curve)*
The most common way to describe the performance of a soaring wing or bird is to draw the glide polar (or polar curve), which is simply the vertical speed versus the horizontal one during gliding (having the same shape as the plot of the lift coefficient versus the drag coefficient). The knowledge of the whole glide polar provides information about *e.g.* the bird's minimum speed (or stall speed) which determines how narrow thermals can be used by the bird, or, from an other point of view, how closely a bird can circle to the strongest, central part of the thermal. Other important data can also be deduced from the polar curve, notably the minimal sinking speed, the best glide ratio and the highest speed.

The polar curve of a soaring bird (white-backed vulture, *Gyps africanus*) was first determined using a camera mounted on a powered sailplane [3]. Photographs were taken with fixed time intervals, and vertical and horizontal speed of the bird with respect to the sailplane was calculated from these pictures. This method had the disadvantage that the exact measurement and subtraction of the air movement from the birds gliding vertical speed was not solved, so the data points were scattered. Although the airplane gave some information about the air movement but not exactly at the bird's location. Later, others determined the glide polar of soaring birds in wind tunnels [4]. Recently, the gigantic extinct volant bird *Argentavis magnificens* (with a mass of 70–72kg, wingspan of 7m, it was the world's largest known flying bird) glide polar has been reconstructed from theoretical calculation based on data obtained from the fossils [5].

Recent studies have shown that variable sweep enlarges the glide polar [6]. Performance (lift and drag coefficients) of the wing of common swift (*Apus apus*) has been measured in a wind tunnel. At low angle of attacks (low horizontal speed) swept wings had lower drag coefficients than extended wings. It was shown that although the best glide performance could be achieved with extended wings, there are flight situations where swept wings are optimal. For example high sweep maximizes high-speed glide performance (at high speeds the glide ratio of a swept wing is higher than the glide ratio of an extended wing at the same velocity) and in addition it was found that swept wings could bear higher loads during fast turns. From the aspect of soaring it seems that extended wings are optimal for turning slowly in thermals, and on a wide range of gliding speeds, but in theory swept wings could be better in weather conditions when thermals are strong and the optimal gliding speed between them is high (see later at 3.1).

*2.2 Dependence of trajectories on wing parameters*
Soaring flight performance and the morphology of three different tropical soaring birds were compared systematically in later studies [7]. Frigatebirds, black vultures and brown pelicans were investigated with ornithodolite from the ground. In one part of that study the dependence of circling radius on wing loading was investigated. The circling path in thermals was reconstructed for a short time period (few circles) and the measured wind was eliminated from the trajectory. The measured circling radius changed between 12m and 18m for the three bird species, growing proportionally to their wing loading. This can be understood if we take a look at the following equation that can be derived from the equality of the forces. The centripetal force and horizontal component of the lift force are in equilibrium during gliding along a helical path (Figure 2):

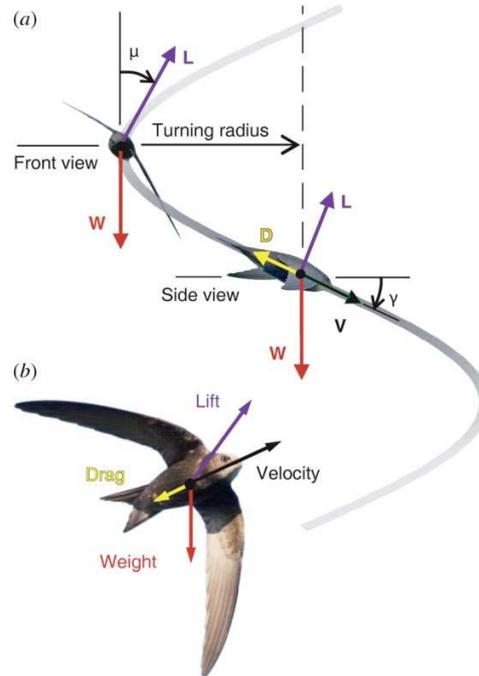

**Figure 2.** Equilibrium gliding of swifts. (*a*) A swift glides on a helical path with constant glide speed (*V*), and a *γ* glide angle. To turn the swift has to bank its body with bank angle ($\mu$), horizontal component of the lift, $L \sin(\mu)$ equals to the centripetal force. (*b*) *L*, lift; *D*, drag; *W*, weight forces acting on a gliding swift. Courtesy of Lentink et al. [6]. Reproduced with permission, © 2007 Nature Publishing Group.

$$r = \frac{W}{S} \frac{2}{\rho g} \frac{\cos^2(\gamma)}{\sin(\mu) C_l},$$

where the following notation was used: *r* - turning radius, *W* - weight, *S* - wing area, *W/S* - wing loading, $\rho$ - air density, $g = 9.806$ m/s$^2$ - standard acceleration due to gravity, $\mu$ - bank angle, $\gamma$ - glide angle, $C_l$ - lift coefficient.

If we assume that the following part of the previous equation

$$B = \frac{\cos^2(\gamma)}{\sin(\mu) C_l}$$

is similar for different bird species [8], we could expect that the circling radius grows linearly with wing loading. This assumption is not valid for very different flying objects. The soaring flight trajectory of the peregrine falcon (*Falco peregrinus*) and the white stork (*Ciconia ciconia*) were recorded recently by GPS and were compared to the flight of paragliders and hang gliders [9]. It was shown that paragliders with very different wing loading (lower than that of storks) fly with roughly the same circling radius as storks, therefore their *B* cannot be the same (Figure 3 (a)).

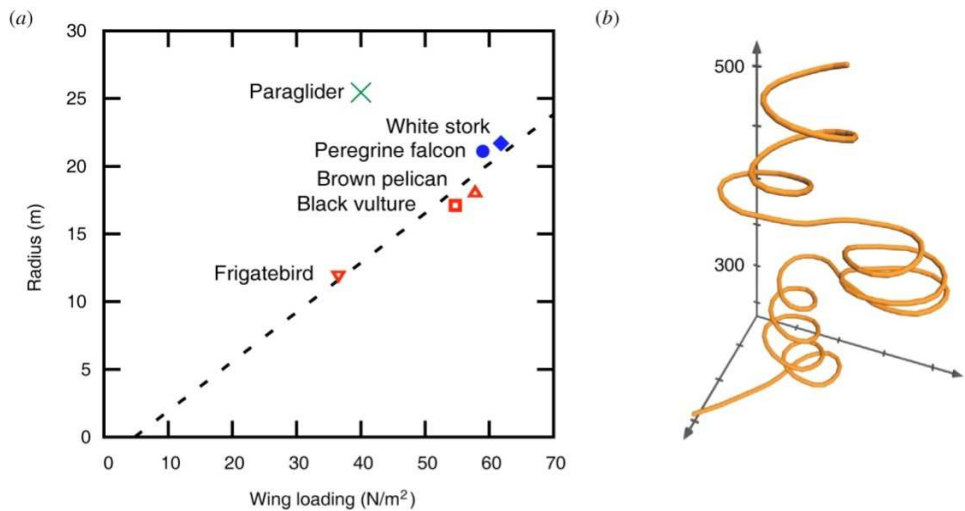

**Figure 3.** (*a*) Circling radius versus wing loading. For three bird species (frigate bird, black vulture, brown pelican; open red marks) Pennycuick found that the main circling radius is linearly proportional to wing loading. Our results for the peregrine falcon and the white stork (solid blue marks) are in agreement with this tendency. The dashed line was fitted to the data on birds. The green X mark shows the measured data for the paraglider. (*b*) GPS logged flight path of a falcon during centring in a thermal. Frequent changes in direction can be seen clearly. Courtesy of Akos et al. [9]. Reproduced with permission, © 2008 PNAS.

The calculated circling radius of the falcon (21.1m) and the stork (21.7m) were also in agreement with the observations and arguments in earlier studies, namely that the mean circling radius is linearly proportional to wing loading.

The differences in aspect ratio (square of the wingspan divided by the area of the wing) have also been discussed in a study on different tropical soaring bird species [7]. High aspect ratio wings have a better lift to drag ratio (which determines how far a gliding object can reach from a given height) and marine soaring birds (albatross, frigatebird) have evolved with relatively high aspect ratio wings. One could think that a high aspect ratio wing is advantageous for every soaring bird, but it seems that birds soaring over land have lower aspect ratios, which might be in connection with their departure technique. While the previously mentioned frigatebird takes off only from elevated perches such as trees, never from level ground or from the water, birds living over land are able to depart from the ground.

*2.3 Making use of the boundary layer versus wing parameters*
Later studies tried to predict which part of the boundary layer is used by the birds [10, 11, 12]. Where can we expect that the trajectories of birds and airplanes cross in a particular weather condition and possibly cause an accident? As it could be expected (since lower wing loading means lower minimum sinking speed), a connection between wing loading and boundary layer usage was found.

It is known from measurements and theoretical calculations that thermals are not equally strong from the ground to the top [13]. The vertical velocity profile depends on the actual weather conditions but usually the middle part is stronger and it is weaker near the ground and at the top. As higher wing loading causes higher minimum sinking speed, birds with higher wing loading are not able to exploit the weaker top and bottom part of the thermal. Another reason could be that the width of a thermal also changes with height (usually the width grows with height and sometimes thermals of different origin join together, forming a wider one). As we have seen earlier, the minimum circling radius of birds also grows with increasing wing loading, therefore the narrow bottom part of the thermal may also not be exploitable for birds with high wing loadings. It is also a possible explanation that those birds that would like to maximize their cross-country speed during flying only use the strongest (usually the middle) part of the thermal.

Use of the boundary layer was measured in a study [10], where the maximal height reached by birds was measured by radar, and the maximal height of the thermals was calculated using the ALPHATERM meteorological model. It was found that the bird with the lowest wing loading (31 $N/m^2$) in this research, the honey buzzard (*Pernis apivorus*), used the biggest part of the boundary layer (91%). The white pelican (*Pelecanus onocrotalus*), which had the highest wing loading (84 $N/m^2$) among the four species investigated, used only 54% of it. White storks (*Ciconia ciconia*) used 69% (wing loading 63$N/m^2$) and lesser spotted eagles (*Aquila pomarina*) used 65% (wing loading 44 $N/m^2$). The boundary layer usage of soaring raptors was also investigated using a Doppler light detection and ranging (lidar) system [14].

One extreme end of the evolution of soaring birds from the aspect of wing loading is the magnificent frigatebird, which has the lowest wing loading among birds (Figure 4 (a)). It was a mystery for a long time how they spend their time at sea until altitude data of the individual frigatebirds have been collected by altimeter (data were sent to the ground by satellite transmitter for analysis)[15]. From the data it has become clear that these birds are continuously on the wing (Figure 4 (b)), day and night during their foraging trip. The curious morphology of these birds is well adapted to the use of very weak thermals that develop over tropical waters in regions affected by trade winds (which blow around the Equator from east to west), even during the night. By circling slowly day and night, high on thermals, these birds very efficiently forage over tropical waters in which prey is scarce. Their average climbing speed in thermals is very low (average climbing rate: 0.40m/s, maximum: 3.3m/s) as the thermals are weak in this area. The price of this very efficient thermalling wing is that these birds have a low speed; the average cross country ground speed of these birds is only around 10km/h.

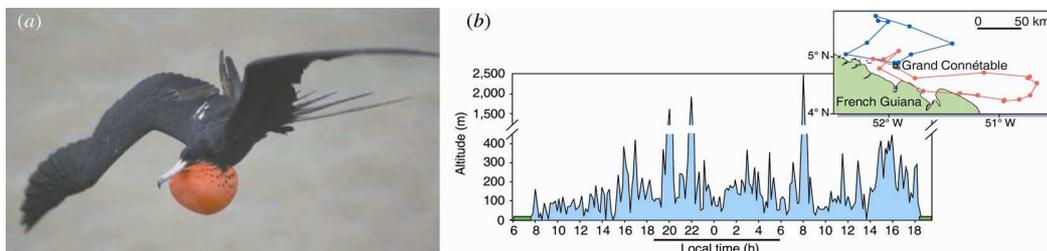

**Figure 4.** (*a*) Magnificent frigatebird (*Fregata magnificens*). (*b*) Altitude data collected during a two day long period, while the bird was foraging and spent some time in its nest (green bars). Black bar indicates hours of darkness (left-bottom panel). Horizontal flight path of two frigatebirds (blue, brooding male; red, incubating female) measured by satellite telemetry during foraging, Green box indicates the colony nest (right upper panel). Courtesy of O. Chastel [15]. Reproduced with permission, © 2003 Nature Publishing Group.

Here we would like to mention that albatrosses have evolved in the other direction of extreme wing loading. Very high wing loading makes birds able to exploit the energy from wind gradients by a different soaring method, called dynamic soaring, and fly great distances over the sea close to the surface [16].

**3. Optimal soaring strategy**

It was a great improvement in the history of gliding when Paul B. MacCready published his theory about soaring flight optimisation [17, 18]. Before the publication, he also won the gliding world championship in 1956.

*3.1 MacCready's speed to fly theory*
The main idea is that sailplane pilots should adjust their gliding speed to the expected thermal climb rate according to their own polar curve. It can be shown that for every expected climb rate there is only one optimal gliding speed on the polar curve.

The main concept of the MacCready theory is the following (assuming still air, where no instantaneous vertical air motion has to be taken into account): the glider tries to optimize its cross country speed, and therefore its goal is to cover a given distance $L_{AB}$ in as short a time as possible. During the flight path from *A* (top of the previous thermal, which is the starting point of a gliding optimisation segment from the flight) to *B* (top of the next thermal), it first glides from *A* to the next thermal and then lifts in it. When it arrives to *B* it has to be at the same height as at the starting point. Thus, it intends to minimize the quantity (time) $L_{AB} [1/v_{xy} - v_z /(v_{xy} v_{climb})]$, where $v_{xy}$, $v_z = p(v_{xy})$ are the gliding horizontal and vertical velocities, $p()$ indicates the polar curve, and $v_{climb}$ denotes the climbing rate in the thermal (see Figure 1 (a)). We find the minimum of $L_{AB} [1/v_{xy} - v_z /(v_{xy} v_{climb})]$ where its derivative is equal to zero and the second derivative is positive. Solving this equation we obtain a relationship between the optimal $v_{xy}$ and $v_{climb}$.

$$\frac{p(v_{xy}) - v_{climb}}{v_{xy}} = \frac{dp(v_{xy})}{dv_{xy}}.$$

This equation expresses that the optimal $v_{xy}$ can be obtained by drawing a line from the point $v_{climb}$ (along the vertical axis) tangent to the $p(v_{xy})$ polar curve and reading the corresponding $v_{xy}$ value (Figure 5). We will refer to MacCready's theory later as "optimal soaring strategy".

*3.2 Comparing bird and human soaring strategies*
The first study that investigated if there is any connection between thermal strength and bird's gliding speed between the thermals studied Marsh harriers in southern Israel by radar and they found positive correlation [19].

More recently, the optimal soaring strategy of birds and competition pilots have been compared [9]. This was the first time that continuous trajectories of soaring birds have been recorded by GPS devices (Figure 1, Figure 3 (b), see also the supplementary movies [20]). Peregrine falcon (*Falco peregrinus*) and white stork (*Ciconia ciconia*), birds specialized for flying great distances, were investigated. Circling radius measurement method of this study has been discussed before, now we give more details about that research. Peregrine falcons use thermals during foraging to soar up to a suitable height from where they can stoop for the prey. Even though they are able to migrate 190km/day with soaring technique [21], scanning a larger area for a shorter time period is more advantageous. Therefore presumably falcons can benefit from higher cross country speed.

For the comparison of the optimal soaring strategy of birds and pilots, GPS flight data of human pilots were also collected. These days, the only way to verify completion of a task at a cross country flying competition is by submitting a GPS track log. At competitions pilots start from one place and have to fly along a route determined by turning points. The objective is to reach the goal as soon as possible, or if it is impossible (*e.g.,* the weather is not good enough to complete the task), to fly as far as possible along the route. A different type of competition is the online contest where pilots fly individually and upload their GPS track logs to the webpage of the contest. This way flights performed at different places and dates can be compared by a pre-defined algorithm, evaluating (amongst others) the length and average cross country speed of the flight. The detailed data about human pilots have been collected from paraglider and hang glider contests, where the pilots are aiming at the highest speed possible to win the competition.

The flight of a falcon has been compared with the flight of top hang glider and paraglider pilots regarding the optimal flight strategy (Figure 5). Hang glider pilots seem to adjust their gliding speed closest to the theoretically optimal. Paraglider pilots use somewhat lower horizontal gliding speed than the optimal (points are more scattered to the left of the optimal curve). This can be interpreted by taking into account that paragliders' glide ratio is worse than that of hang gliders, so they choose a lower speed to minimize the risk of not reaching the next thermal before landing. In addition, paragliders have lower stability at higher speeds, so in some situations, pilots do not apply the maximum speed for safety reasons. Falcons seem to adjust their flight to the actual weather conditions according to the MacCready theory as well. Thus, as it happens, evolving flight strategies of birds and human calculations lead to virtually the same outcome. To answer the original question, it seems that those bird species which can benefit from it, instinctively apply the "speed to fly theory" that was originally calculated and described by Paul MacCready, and since then is widely used by pilots in thermal soaring.

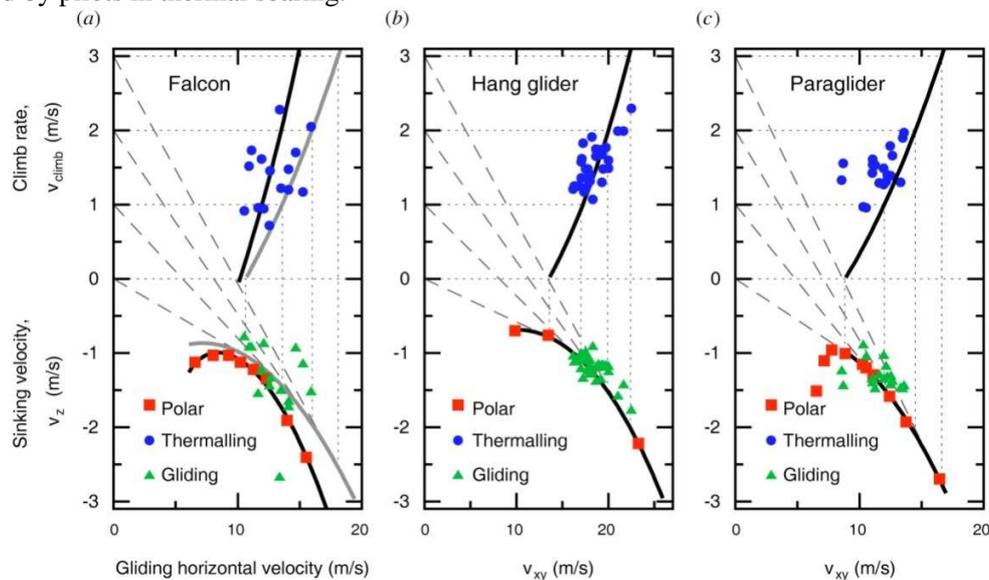

**Figure 5.** Soaring strategy plot of various flyers. Squares denote data points measured in the wind tunnel that the polar curve was fitted to (solid line at the lower part). From the polar curve the optimal strategy curve (solid line in the upper part) is calculated by using the MacCready theory. The dashed and dotted lines illustrate this process. The dashed lines show tangent lines to the polar curve from points corresponding to different climb rates represented on the positive part of the $y$ axis. Dotted lines indicate for each dashed line the corresponding climb rate in the thermal lift and the optimal horizontal velocity for gliding between the thermals. Circles denote the averaged climb rate versus the measured averaged gliding horizontal velocity (one circle for one flight). Triangles indicate the averaged gliding horizontal speed as a function of the sinking velocity. (*a*) Peregrine Falcon. Gray curves show the effective polar curve (including flapping flight during the gliding parts) and the corresponding optimal strategy curve. (*b*) Flight data for hang glider and (*c*) two paraglider pilots achieving the third and the first two places in the online world contest (OLC), respectively. Courtesy of Akos et al. [9]. Reproduced with permission, © 2008 PNAS.

**4. Soaring UAVs : advances and challenges**

Small UAVs have payload restrictions and can carry fuel/energy for only a short amount of time. However, several researchers have already demonstrated in simulations and real applications that not just birds and manned gliders, but also UAVs are able to exploit updrafts to increase their endurance and save energy. For example, theoretical calculations by NASA are very promising as they showed that an UAV with a nominal endurance of 2 hours could fly a maximum of 14 hours using updrafts in good weather conditions [22].

In this fourth part we will provide an overview of existing research on different thermal exploitation strategies of UAVs both in simulations (and challenges of simulations like creation of a realistic thermal model and simple but effective flight control strategies) and applications on real platforms. The objective of these investigations is to find the center of the thermal in a circling maneuver and maximize the net lift over the searching path inside the thermal. Thermal exploitation methods have to work effectively in different thermals, like wide, narrow, etc.

In the final section we will demonstrate how the addition of noise (that imitates turbulence) to the existing thermal models (and thus approximating real thermals more accurately) changes the effectiveness of thermal exploitation strategies. Two different, GPS independent strategies will be compared in a thermal model with perturbation: the simplest commonly used strategy and a new one that was inspired by the falcon's flight.

*4.1 Review of previous simulations*
John Wharington was the first who proposed an autonomous soaring simulation for UAVs in 1998 [23]. In this work he gave a framework of updraft modeling for UAV soaring strategy simulations.

Thermals are turbulent airflows, so accurate numerical simulation of the airflow in the thermal updraft would be very difficult and would require a lot of computation power. Instead of this complicated simulation he used a very simplified thermal model, based on measurements and theoretical calculations. In this model the thermal is a circle or ellipsoid shaped vertical tube, usually with Gaussian vertical velocity distribution. In later simulations others used very similar, quadratic function shaped [24] or radius dependent [13] thermal vertical velocity distributions.

First he tested a strategy that is well known among glider pilots, based on the rules of the famous competition glider pilot, Helmut Reichmann [25]. These rules are the following:
1. If climb improves, decrease bank angle
2. If climb deteriorates, increase bank angle
3. If climb remains constant, keep the bank angle constant

A simple strategy based on the rules above worked quite efficiently in simulations. But as Wharington demonstrated in his work, this rough method is oversimplified and other methods that account for the thermal profile and the influence of variometer errors could be more effective. He developed a guidance algorithm using reinforcement learning and a neural-based thermal center locator for the optimal autonomous exploitation of the thermals. The thermal center locator part of the algorithm had the same function in the simulation as the following centring rule that pilots apply during thermalling in addition to the Reichmann rules. During the last completed circle they memorize their heading when the lift was the strongest. Then they try to move the circle towards the lift (away from the sink which is usually on the other side). This method is based on their sense of orientation (practically by noting a landmark over the plane's nose where the lift was strongest) [26]. However, Wharington's algorithm with the neural-based thermal center locator was too time consuming for real time on-board applications.

Stephane Doncieux and his colleagues developed a soaring strategy by using evolutionary algorithm to optimize the connection weights in a neural network [27]. Altitude gain during the allotted evaluation period was used as fitness function. The input parameters of the neural network were the vertical velocity, roll and pitch angle of the glider. The outputs of the network controlled the elevator and the rudder of the plane. In simulations this strategy was successful for ideal thermals, modeled similarly to those proposed by Wharington [22].

*4.2 Realisations of Autonomous soaring*

Two different successful thermal exploitation strategies have been programmed into UAV in reality. Both strategies' centring method is fully or partially based on position information-coupled updraft data collected in the past, which could be the key feature of their success. Simpler strategies that rely on instantaneous updraft measurements have been tried out in reality [28], but the authors suggest themselves that a more elaborate strategy is most probably necessary to track and stay near the center of a thermal.

The first successful soaring UAV was developed by NASA during the Autonomous Soaring Project at Dryden Flight Research Center [29]. Michael Allen and a team of engineers programmed a small UAV (a 4.27m wingspan, 6.8kg motor-glider called Cloud Swift, see Figure 6) to detect whether it is in an updraft and use that updraft by circling. During the project the UAV flew 17 times, gained an average altitude of 173m in 23 updrafts, and ascended 844m in one strong thermal. In one flight of this project the UAV added 60 minutes to its endurance by soaring autonomously.

Most recently even more successful thermal locating and guidance algorithms have been designed and implemented at the North Carolina State University and the U.S. Naval Research Laboratory by Daniel J. Edwards [30, 31]. The UAV in this project took part in the Cal Valley Soaring Race in May of 2009 where it beated humans in a head-to-head cross-country competition and covered over 113km. In another flight it stayed aloft for over 5.3hr.

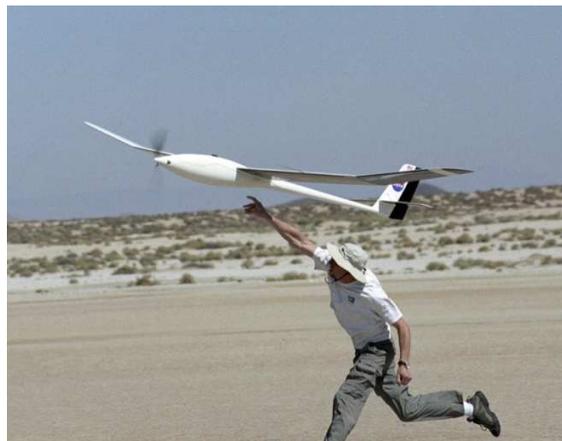

**Figure 6.** Allen from NASA Dryden Research Center launches a soaring UAV (Cloud Swift) by hand. Courtesy of Carla Thomas, NASA Photo. Reproduced with permission, © 2005 NASA Dryden Flight Research Center Photo Collection.

In NASA's Autonomous Soaring Project the algorithms were run in real-time on board as part of the autopilot hardware of the Cloud Swift. First the updraft velocity was calculated from the aircraft's motion to determine if the plane was in an updraft and to estimate the location and strength of the thermal. It is easy to see that the aircraft's growing altitude (and the growing potential energy, $E_{PE}$) is connected to the updraft velocity. As the kinetic energy ($E_{KE}$) of an aircraft can be converted into potential energy (imagine a fast moving aircraft turning upwards and slowing down), the kinetic energy of the plane should also be taken into account. The total energy ($E_{total}$) of the aircraft is the sum of instantaneous potential and kinetic energy values, calculated as follows:

$$E_{PE} = mgh,$$

$$E_{KE} = \frac{1}{2}mV^2,$$

$$E_{total} = E_{PE} + E_{KE},$$

where *m* denotes the mass, *g* refers to the acceleration due to gravity, *h* is the altitude, and *V* indicates the velocity.

The derivative of the plane's total energy ($\dot{E}$) gives the total energy rate that is inherently coupled to the updraft velocity (to give the net updraft velocity the aircraft sink rate had to be added to the measured total energy rate). The second derivative of the plane's total energy (energy acceleration) gives information about changes of the updraft velocity used in Allen's study as one input parameter for the controller. The direct effect of the energy acceleration on the control can be interpreted as the application of Reichmann rules described above.

Additionally, to find the core of the thermal more reliably, a Gaussian thermal model (described above) was also fitted to the last two circles of the flight path logged by the GPS in the following way. Wind was calculated from the drift of the thermal and was eliminated from the trajectory to get almost neat circles (in other words data was transformed from the earth coordinate system to a coordinate system that was drifting with the thermal). The location of the thermal's center was calculated from an updraft weighted average using the expression:

$$\mathbf{P}_{th} = \left[ \frac{\sum lat \cdot \dot{E}^2}{\sum \dot{E}^2} \quad \frac{\sum lon \cdot \dot{E}^2}{\sum \dot{E}^2} \right]$$

Here $\mathbf{P}_{th}=[lat_{ud}\ lon_{ud}]$ is the calculated location of the center of the updraft, *lat* and *lon* are the coordinates of the GPS position and $\dot{E}$ is the corresponding calculated total energy rate in that position. Using this estimated updraft center there remain two parameters that are needed to describe the thermal, the thermal strength, and the characteristic radius. The maximum velocity of the thermal was estimated simply by adding 10% to the measured maximum updraft. The radius of the thermal was calculated by using an iterative fit to an assumed Gaussian shaped thermal vertical velocity profile. Based on the location and radius information of the fitted thermal model three other input parameters were determined for the controller (position error, velocity error, and the steady-state turn rate) in addition to the energy acceleration. These four input parameters were tuned to give an appropriate circling turn-rate command as an output parameter for the autopilot.

The soaring strategy of the UAV programmed by Daniel J. Edwards has two main differences compared to Allen's implementation that was described in detail above: the implementation was off-board and included an improved method for thermal location estimation. The on board autopilot received the desired high-level commands with 1Hz via radio connection from a distant laptop.

Allen's thermal center location algorithm was used in the research by Daniel J. Edwards as a starting point, but since higher computational power was available due to the off-board system than before, Wharington's (see above) neural network based method was also integrated into the final solution. The main idea was that the center of the thermal has to be searched on a grid of nodes in the vicinity of the initial thermal center (calculated by Allen's modified center of mass method). The positions and updraft data of the flight path was calculated and logged. Wind was calculated by averaging the wind data that was provided by the autopilot, after which drift correction and the initial thermal center was calculated similarly to the previous work. The usual Gaussian thermal model was used. A standard linear regression method was applied (after transforming the thermal model to log space) to calculate how well the measured data correlated with the possible thermals with different centers (in the nodes around the calculated initial thermal center). The thermal allocated to the node with the best correlation was chosen as the thermal center. Figure 7 shows the difference between Allen's and Edwards' thermal center location methods.

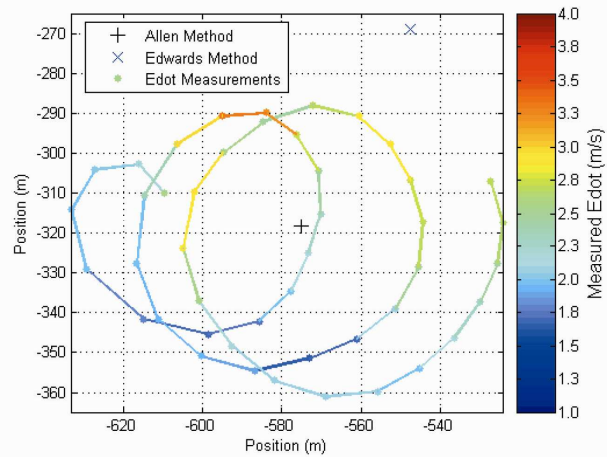

**Figure 7.** Comparison of the two different thermal center estimation methods. The figure shows the GPS logged flight path of the UAV and the corresponding colour coded $\dot{E}$ values. The main difference is that the newer method (Edwards') makes it possible to locate the thermal center outside of the area covered by the measured data. Allen's method estimates the thermal center always somewhere inside the area of measurements, while Edwards' method places the thermal center into a location that would yield concentric rings of similar $\dot{E}$ values. Courtesy of Daniel J. Edwards [30]. Reproduced with permission, © 2008 American Institute of Aeronautics and Astronautics.

Gliding speed has also been optimised in this research using MacCreadys' speed to fly theory, to increase cross-country performance.

*4.3 The idea of a soaring flock of UAVs*
The idea of a soaring flock of UAVs was first proposed by John Wharington. A soaring flock (several UAVs together) would be even more efficient in the exploitation of the natural energy of thermals than a single UAV. The flock could cover larger areas together, searching for the best thermal, and with radio connection they could share their knowledge with each other. With this strategy they could minimize the time spent circling in thermals by choosing that flock member's thermal that was found to be the strongest. Cross country soaring competition pilots widely apply this strategy. They are continuously watching each other and change their thermal for the other pilot's one (or a bird's one), if the other seems to gain height faster. Additionally, a combination of UAVs with different aerodynamic properties in one flock would have the advantage that the „faster" ones with better lift to drag ratio could search for thermals and objects over a large area, whereas the „slower" ones with the ability of exploiting weak updrafts could stay more efficiently in one place if needed and directly go to the thermal or observation point which has been discovered by the "faster" ones.

*4.4 Simulations of GPS independent thermal exploitation strategies in turbulent thermals*
A flock of UAVs is theoretically capable of establishing wireless networks between multiple ground-users relying only on local communication with immediate neighbours and sensors that provide heading, speed, altitude and angular velocities instead of GPS [32]. Such UAV networks can play an important role in disaster mitigation. Projects in which UAVs are fully independent, so they do not rely on any external positioning system have motivated the ongoing cooperating research project of Laboratory of Intelligent

Systems, École Polytechnique Fédérale de Lausanne (LIS, EPFL) and Department of Biological Physics, Eötvös University (ELTE) to develop GPS free soaring guidance methods for UAVs to exploit thermals.

As we described above, both thermal exploitation strategies implemented into real applications used position information-coupled updraft data collected in the past (the plane was "mapping the thermal") for the thermal center calculation. But as we have seen in the simulation part, very simple strategies exist which do not rely on previously collected data and therefore they do not use position information either.

However, all of the above simulations [13, 23, 24] made use of very simple thermal models. We aim at demonstrating how the efficiency of the simplest strategy, the so-called "Reichmann rules" (see the details of the strategy above) changes depending on the thermal model. Therefore, we modified the thermal model [24] to include turbulence in the form of an additional 3D noise. Random values selected from the interval [-1;1] with uniform distribution were assigned to a three-dimensional set of mesh points and smoothed with a Gaussian filter, see Figure 8 in an analogy to the way turbulence was accounted for in a dynamic soaring simulation [33].

The velocity profile of the thermal ($\mathbf{w} = (w_x, w_y, w_z)$ the local air velocity vector at position $(x, y, z)$) in the vertical direction was defined as a quadratic function:

$$w_x(x,y,z) = 0, \quad w_y(x,y,z) = 0, \quad w_z(x,y,z) = \max(w^{max} \cdot (1 - \frac{(x-x_0)^2 + (y-y_0)^2}{r^2}), 0),$$

where $w^{max}$ the maximal velocity value at the center of the thermal which is at position $(x_0, y_0)$ and $r$ is the radius of the thermal. The velocity of the airflow with noise ($\mathbf{w}^* = (w_x^*, w_y^*, w_z^*)$) was calculated as:

$$w_x^*(\mathbf{r}) = w_z(\mathbf{r}) \cdot \eta \cdot \xi_x(\mathbf{r}),$$
$$w_y^*(\mathbf{r}) = w_z(\mathbf{r}) \cdot \eta \cdot \xi_y(\mathbf{r}),$$
$$w_z^*(\mathbf{r}) = w_z(\mathbf{r}) \cdot (1 + \eta \cdot \xi_z(\mathbf{r})),$$

where $\eta$ defines the strength of the local perturbation vector $\xi(\mathbf{r}) = (\xi_x(\mathbf{r}), \xi_y(\mathbf{r}), \xi_z(\mathbf{r}))$ at position $\mathbf{r} = (x, y, z)$. For $\xi(\mathbf{r})$ random numbers were generated from the interval [-1; 1] with uniform distribution in a set of mesh points of a finite cubic grid of 50 x 50 x 100. A Gaussian averaging (three-dimensional Gaussian function with $\sigma=2$, thus the length scale of the perturbation of the air flow was 2 m) was used to smooth the noise (with periodic boundary condition). The local perturbation $\xi(\mathbf{r})$ was calculated as the weighted average of the values in the eight neighbouring grid points, where the weights were defined as the inverse of the distances. The horizontal components of the velocity of the airflow with noise have mean values of zero and their strengths are proportional of the local strength of the updraft.

It should be noted that to prove the efficiency of a strategy more realistic simulation with a more sophisticated environmental model is needed such as the Dryden Gust Model [34]. Here we want to show the effect of wind turbulence on the performance of a strategy in a simple way, and investigation of a more realistic model will be a subject of a further study.

Matlab simulation of the Sky-Sailor project [24, 35, 36, 37] (wingspan 3,2 m, weight 2.4 kg, wing area 0.776 m$^2$), was used as a basis of our simulation. Figure 9 shows the trajectories of the plane in thermal without noise and with noise. From the plots it can be seen how the noise of the thermal misleads the strategy that easily found the thermal center in the simple thermal model. The height gained during the same amount of time both as the function of turbulence and the thermal radius were studied. The effectiveness of the strategy was found to decrease for growing noise and decreasing radius (see later). These results inspired us to try out another very simple GPS independent method that relies not just on actual measured updraft data but also memorizes and uses previous updraft information (and therefore may be less effected by thermal turbulence).

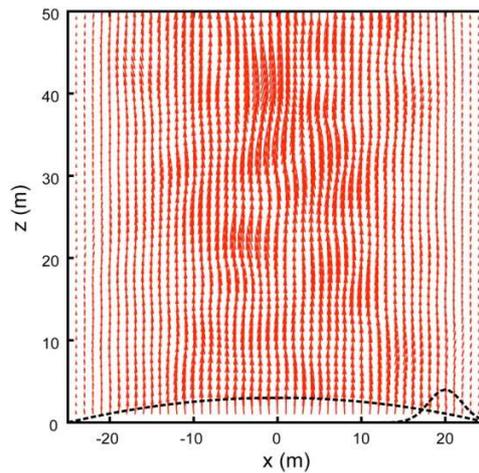

**Figure 8.** Two-dimensional section of the thermal model ($r=25$ m), with quadratic vertical velocity distribution and additional random 3D noise for $\eta=3$. The quadratic vertical velocity distribution of the thermal is noted with black line. The shape of the smoothing Gaussian function for $\sigma=2$ m can be seen in the lower right corner.

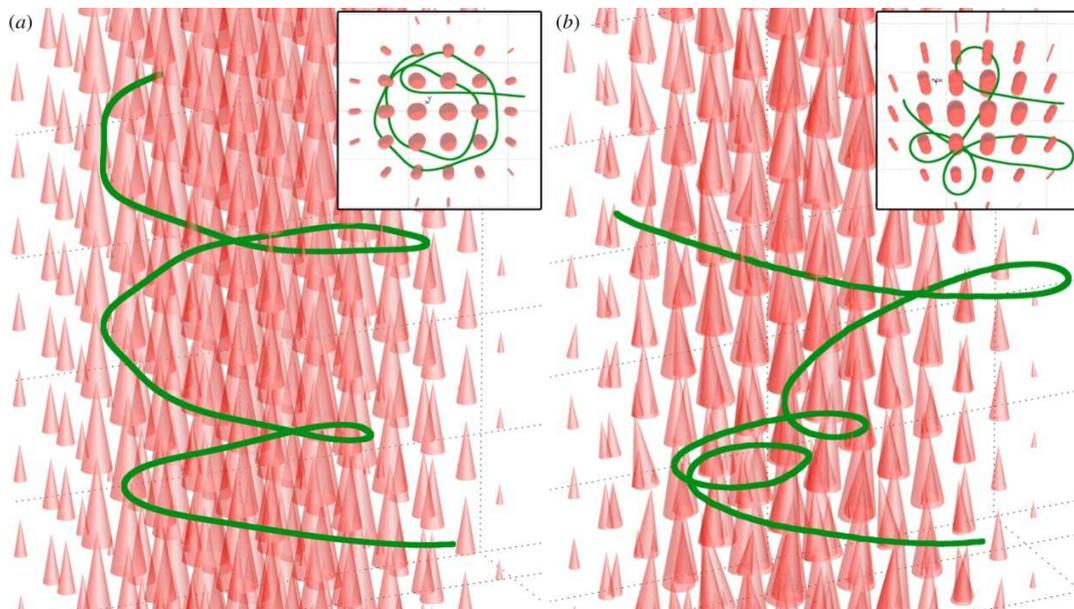

**Figure 9.** Analysis of RR thermal exploitation strategy in a thermal model without and with perturbation. Trajectory (green line) of the thermalling simulated UAV in a thermal tube ($r=200$ m) (a) without perturbation ($\eta=0$) and (b) with perturbation ($\eta=2$). Red cones indicate the local air velocity. Insets in the upper right corners show the top view of the same trajectories.

Thermal exploitation strategy of birds has not been systematically compared to those used by pilots or UAVs, so it is not known if birds are more (or less) successful in soaring, and to judge this question further research would be required. But the concept that thermal exploitation strategy of birds is very efficient, perhaps even more efficient than pilots' practice (which can also be very complex and hard to be described fully, see above in 4.1) is supported by the observation of pilots that birds usually circle in the core of the thermal. Typically, if a pilot sees a bird nearby in the same thermal, usually it is worthwhile to move his circle towards to the bird's flight path. It is not clear how birds solve this task. It is possible that they apply some method similar to the Reichmann rules (above) and they may also be actively mapping the thermal somehow as in the strategies described above. But it seems less likely that birds could reconstruct their flight path in the earth reference frame for mapping the thermal as GPS could do and compare it to an imagined drifting thermal, especially when they are at high altitude.

GPS logged flight path (Figure 3 (b) and supplementary movies [20]) of the peregrine falcon, from where it can be seen that falcon changes his circling direction frequently and thus shifts the center of circling during thermalling, motivated us to try out a very simple direction changing (DC) strategy (Figure 10) based only on heading and updraft information of the UAV (requiring only magnetometer and pressure sensor) and compare it to Reichmann rules (Figure 11). Direction change is a technique that is sometimes also used by pilots during centring and makes it possible to shift the center of the circling path (by two radii) while flying on a smoothly changing trajectory. This strategy is similar to the strategies implemented in real applications from the point that the thermal center calculation is based on updraft information collected in the past (the plane is "mapping the thermal"), in contrast to the RR strategy where instantaneous updraft information is used. The controller works in the following way: the heading and the corresponding updraft value is logged with fixed sampling rate for one complete circle after the circling direction of the plane was changed. The heading vectors $\mathbf{n}(\varphi)$ are weighted with the third power of the updraft $\dot{E}$, and the sum of these vectors over one complete circle ($2\pi$ change in heading direction $\varphi$) gives the direction of the expected thermal center $\mathbf{D}_{th}$ :

$$\mathbf{D}_{th} = \sum_{\varphi=0}^{2\pi} \mathbf{n}(\varphi) \cdot \dot{E}^3 .$$

When the plane's heading gets close to $\mathbf{D}_{th}$ during the next circle, the plane changes circling direction. The disadvantage of using only magnetometer data is that the plane has to fly in neat circles to be able to decide which direction it should shift its circle, and has to fly at least one complete circle. The advantage compared to GPS based methods is that this may require reduced computation power as the wind elimination step could be skipped (since the magnetometer is also drifting with the thermal). For simplicity the plane flies with constant circling radius in this simulation and is forced to change direction after every complete circle when the plane's heading is close to $\mathbf{D}_{th}$ (this can be improved by implementing variable circling radius, and the possibility to only shift the center of the circle with a fraction of the radius towards the direction of the thermal center instead of a direction change, into the strategy). As we can see in Figure 1 (b) the DC strategy was more effective and stable (smaller deviation) in thermals with perturbation ($\eta=2$) for smaller radius values (100-200m), but it can be also seen that RR strategy is the winner for bigger thermals (250-300m) where searching the thermal center with the DC method starts to be less effective. Figure 11 (a) shows performance of the two strategies as the function of noise. From $\eta=0$ (no perturbation at all) to $\eta=2$ DC seems a little bit more effective, but for high $\eta$ values (2-3.5) the ratio of crash is higher for the DC strategy which can be attributed to the fact that changing the direction of the plane is more risky in very turbulent thermals.

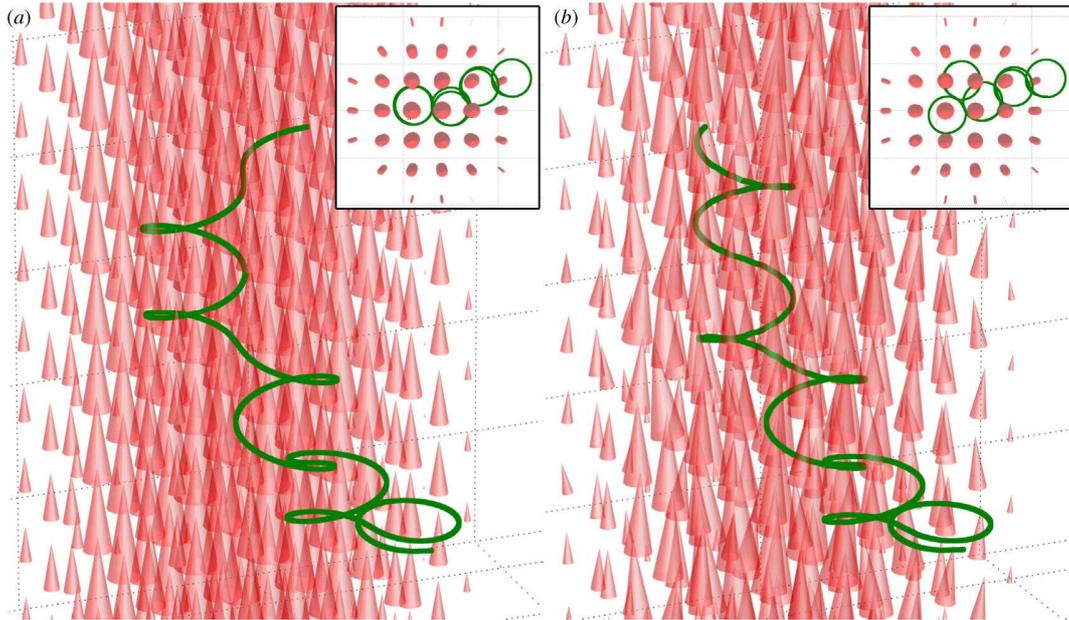

**Figure 10.** Analysis of DC thermal exploitation strategy in a thermal model without (a) and with (b) perturbation. Parameters are identical with Figure 9.

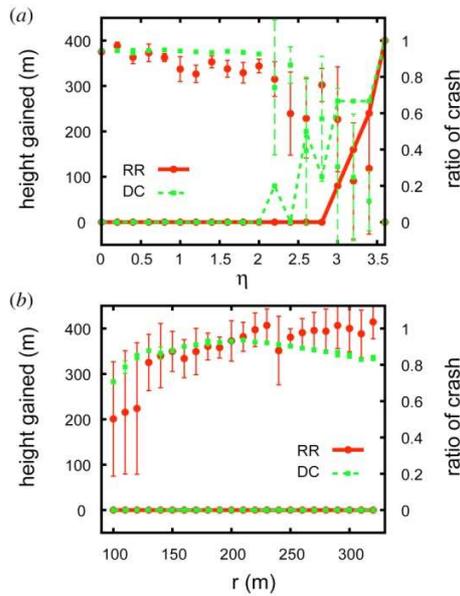

**Figure 11.** Comparison of the simplest thermal exploitation strategy, the so-called "Reichmann rules" (RR, red dots and solid curve), and the simple direction changing strategy (DC, green squares and dashed line) in thermal model with perturbation. For each data point the simulation has been run for 5 randomly generated turbulent thermals. The green squares (DC strategy, see details in the text) and red dots (RR strategy) note the mean value of the 5 simulation runs in different thermals, and the standard deviation is also shown. The lines at the lower part indicate the ratio of crash. (a) Height gained during 200 s long simulation as the function of noise ($\eta$) for $r=200$. (b) Height gained during 200 s long simulation as the function of the thermal radius ($r$) for $\eta=2$.

## Conclusion

Thermal soaring is a form of flight that saves much energy for birds, pilots and UAVs by exploiting nature's energy present in the form of thermals. In the first part we reviewed research on soaring birds in the context of their gliding performance and the dependence of this performance on the ways the birds' wings can be swept. We also discussed the question of the relation of the boundary layer usage and the soaring flight trajectory parameters (like circling radius) to the wing parameters. Recent research investigating the GPS logged flight path of peregrine falcon showed that falcons do apply the "speed to fly" theory, well known and widely applied by gliding pilots to optimise their cross country soaring flight.

In the next part, an overview on state-of-the-art simulations and real applications of thermal exploiting strategies of UAVs were described. The two successful experimental projects on soaring UAVs, NASA's Autonomous Soaring Project by Michel Allen and a project by North Carolina State University and the U.S. Naval Research Laboratory by Daniel J. Edwards were described in detail and compared.

We pointed out that the thermal model commonly used in simulations is oversimplified compared to real thermals, and a systematic analysis of thermal exploitation strategies in simulations could be useful for searching their weaknesses. A more sophisticated thermal model including a 3D noise as imitation of turbulence was introduced. It was presented how the effectiveness of the simplest thermal exploitation strategy, the so-called Reichmann rules decreases with growing turbulence and decreasing thermal radius.

Finally, possible future work on soaring birds aimed at further improving the thermal exploitation capability of UAVs, and ideas on GPS independent thermal exploitation strategies were discussed. One simple strategy (inspired by GPS logged trajectory of falcons), in which circling direction changes are present, was simulated in Matlab. This direction changing strategy, which actively "maps" the thermal by using past updraft information coupled to heading data, was compared to the applied Reichmann rules which rely on instantaneous updraft measurement. For different radius and turbulence values different strategies were found to be better. Therefore, we suggest that a combination of these two or similar strategies could be a possible solution for achieving GPS independent thermalling UAVs in the future.

Anatomical features of birds, for example feathers, slots at the wing tip between primary feathers [38] or tails that can be moved separately from the wings [39] may also play an important role in their enviable thermal exploiting talent. Unfortunately the above favourable features are not available for manmade wings today. Application of research on morphing wings could also be advantageous in the design of soaring UAVs [6, 40]. Further research on the role of the unique features of a bird wing in thermal exploitation can be useful to improve the wings of soaring UAVs and the thermal guidance methods for optimal soaring.

## Acknowledgements
The research on the GPS free guidance method was initiated while one of the authors (Zs. A) was hosted by LIS, EPFL. We would like to thank D. Floreano for useful discussions. We thank D. Ábel and E. Méhes for their useful comments and grammatical corrections. This research was partially supported by the FP7 ERC Grant COLLMOT.